# *In Silico* Analysis of Tandem Repeats in GIF of Gastric Parietal Cells

Sim-Hui Tee

Multimedia University
Persiaran Multimedia, 63100 Cyberjaya, Selangor, Malaysia
shtee@mmu.edu.my

*Abstract-* **Tandem repeats are ubiquitous in the genome of organisms and their mutated forms play a vital role in pathogenesis. In this study, tandem repeats in Gastric Intrinsic Factor (GIF) of gastric parietal cells have been investigated using an *in silico* approach. Six types of the nucleotide tandem repeat motifs have been investigated, including mono-, di-, tri-, tetra-, penta- and hexanucleotide. The distribution of the repeat motifs in the gene was analyzed. The results of this study provide an insight into the biomolecular mechanisms and pathogenesis implicated by the GIF of gastric parietal cells. Based on the findings of the tandem repeats in GIF of gastric parietal cells, therapeutic strategies and disease markers may be developed accordingly by the biomedical scientists.**

*Keywords- Bioinformatics; Stomach epithelium cells, Parietal cells, tandem repeats, in silico, gastric intrinsic factor*

## I. INTRODUCTION

Parietal cells are the stomach epithelium cells that secrete hydrochloric acid and intrinsic factors in response to the receptor binding of histamine, gastrin, and acetylcholine [1-2]. Acid secretion by parietal cells was observed to be mediated by $H^+/K^+$-ATPase [3], either via histaminergic pathway or cholinergic pathway [4]. Gastrin and the increased intracellular cAMP levels are required to mobilize histamine in order to induce the acid secretion by parietal cells [5]. The high expression of gastrin receptors in parietal cells have been established experimentally [6]. Aberration in stomach such as chronic infection of gastric mucosa by antigens can develop chronic inflammation in gastric mucosa, with the corollary of the diminished level of gastric acid secretion in stomach [7]. The exacerbated inflammation in stomach would decrease the number of active parietal cells and further reduce the level of secreted gastric acid, which would result in gastric ulcer [7].

Because parietal cells play an essential role in the mediation of mucosal proliferation in gastric inflammation [8], it is vital to understand the nature of the nucleotide sequences of this cell type. One of the critical aspects to be looking into is the simple sequence tandem repeats in the genes, as these tandem repeats always are associated with diseases when mutated [9]. Tandem repeat expansion mutations are implicated in about 20 diseases in human, with complicated underlying mechanisms that are not fully unveiled [10]. However, progresses have been made in certain type of tandem repeat expansion, such as trinucleotide expansion [11]. It was observed that the changes in trinucleotide length have substantial impacts, albeit mammals have evolved mechanisms that could resist the deleterious effects incurred by rapid trinucleotide expansion [11]. The high mutation rate in tandem repeats is primarily due to the error in DNA replication that leads to the changes in the number of repeat units [12].

Despite the observed pathogenesis of tandem repeat expansion, it was reported that tandem repeats are ubiquitous in all organisms, and they could occur on any locus on a gene [9]. Tandem repeats have the tendency to accumulate in organisms with larger genomes, such as that of mammals [14]. In addition, it is also known that variable tandem repeats in the genome of an organism can accelerate the evolution of coding and regulatory gene sequences [12]. The information obtained from the investigation of tandem repeats has widespread application in various fields such as medicine, forensic science, and population genetics [13]. Notably, the identification of tandem repeats has special function in genetic studies, such as gene mapping [15].

In this study, tandem repeats in Gastric Intrinsic Factor (GIF) of gastric parietal cells have been investigated using an *in silico* approach. GIF is required for the absorption of vitamin B12, without which could result in megaloblastic anemia, atrophic gastritis, and pernicious anemia. *In silico* approach appears as an efficient approach that has been widely used in the investigation of biomedical phenomena. A wide range of in silico techniques and tools has been established, including algorithms [16-23], databases [24-29], web servers [30-34], and modellings [35-42], which make possible the efficient processing and analysis of large volume of biomedical data. The results of this study provide an insight into the biomolecular mechanisms and pathogenesis implicated by the gastric parietal cells. Based on the finding of the tandem repeats in GIF of gastric parietal cells, therapeutics strategies and disease markers may be developed accordingly by the biomedical practitioners.

## II. METHODS

The nucleotide sequences of GIF of gastric parietal cells have been retrieved from NCBI GenBank. An alignment was done for all the perfect repeat units. A combinatorial-based *mreps* algorithm [43] was used to find the repeated sequence: (1) Given an error threshold $k$, the maximal $k$-mismatch runs of tandem repeats were sought. (2) The left and right edges of the tandem repeats were trimmed. (3) The best period of each repeat was computed. (4) The statistically expected repeats were identified. (5) The tandem repeats with the same period $p$ overlapping by at least $2p$ were merged.

To identify spatial motifs in GIF (for interacting sequences), the following formula was used [44]:

$$P(X,Y) = \frac{f_{obs}(X,Y)}{IE[f(X,Y)]} \qquad (1)$$

where $P(X,Y)$ is the propensity of the residue interacting pair X-Y; $f_{obs}(X,Y)$ is the actual count of X-Y contacts in the interacting sequences; and $IE[f(X,Y)]$ is the expected count.

Tandem repeat motif occurrence was determined using hierarchical gene-set genetics based algorithm [45]. Given $S^{ij}$ which is the subsequence of length $W$ at position $j$ in a sequence $i$. Let $a$ be the symbol that occurs at a position $k$ of either motif or non-motif; let the position $k$ be $1 \le k \le W$. Let $\theta^M$ and $\theta^B$ model the motif and non-motif positions in a sequence, respectively. The conditional probabilities that $S^{ij}$ is found using the motif model and background (non-motif) model are computed as such [45], respectively:

$$P_M(S^{ij}) = \prod_{k=1}^{W} \prod_{a=1}^{L} (\theta_{ak}^M)^{I(S_{i,j+k-1}=a)} \quad (2)$$

$$P_B(S^{ij}) = \prod_{k=1}^{W} \prod_{a=1}^{L} (\theta_{a0}^B)^{I(S_{i,j+k-1}=a)} \quad (3)$$

The motif model $\Theta$ is given in the following equation [45]:

$$\Theta = \{\Theta^B, \Theta^M\} = \begin{bmatrix} \theta_{-,0}^B & \theta_{-,1}^M & \theta_{-,2}^M & \dots & \theta_{-,w}^M \\ \theta_{a1,0}^B & \theta_{a1,1}^M & \theta_{a1,2}^M & \dots & \theta_{a1,w}^M \\ \theta_{a2,0}^B & \theta_{a2,1}^M & \theta_{a2,2}^M & \dots & \theta_{a2,w}^M \\ \dots & \dots & \dots & & \\ \theta_{aj,0}^B & \theta_{aj,1}^M & \theta_{aj,2}^M & \dots & \theta_{aj,w}^M \end{bmatrix} \quad (4)$$

Let $\lambda$ be the prior probability of motif occurrence in the gene sequences. The motif occurrence probability $Z$ at position $j$ in sequence $i$ is derived from [45]:

$$Z_{ij} = \frac{\lambda P_M S^{ij}}{(\lambda P_M S^{ij}) + (1-\lambda) P_B S^{ij}} \quad (5)$$

We take $S^{ij}$ to be a motif encounter when the following is fulfilled [45]:

$$\log(P_M(S^{ij}) / P_B(S^{ij})) \ge \log[(1-\lambda)/\lambda] \quad (6)$$

In this research, six types of tandem repeat motifs were analyzed (mono-, di-, tri-, tetra-, penta-, and hexanucleotide). Relative frequency was used as a measure to analyze the total repeat per kilobase in the nucleotide sequence of GIF. A triplet classification system [46] was used to categorize the trinucleotide tandem repeats.

## III. RESULTS AND DISCUSSION

GIF of gastric parietal cells is 94270 base pair (bp) in length, which is located on chromosome 11. The composition of GIF is: A = 27.02%, T = 26.66%, G = 23.47%, and C = 22.86%. A total of 4919 perfect tandem repeats were found and retrieved from GIF gene sequence. The distribution of tandem repeats of GIF is shown in Table 1.

Table 1. Distribution of tandem repeats of GIF

| Repeat motif | No. of occurrence | Relative frequency |
|---|---|---|
| Mononucleotide: | | |
|   A | 100 | 1.06 |
|   T | 84 | 0.89 |
|   C | 15 | 0.16 |
|   G | 13 | 0.14 |
| Total: | 212 | 2.25 |
| | | |
| Dinucleotide: | | |
|   AT/TA | 548 | 5.81 |
|   AC/CA | 517 | 5.48 |
|   AG/GA | 712 | 7.55 |
|   CG/GC | 191 | 2.03 |
|   GT/TG | 555 | 5.89 |
|   CT/TC | 705 | 7.48 |
| Total: | 3228 | 34.24 |
| | | |
| Trinucleotide | 1048 | 11.12 |
| | | |
| Tetranucleotide | 219 | 2.32 |
| | | |
| Pentanucleotide | 40 | 0.42 |
| | | |
| Hexanucleotide | 8 | 0.08 |

From Table 1, it is clear that dinucleotide is the most abundant tandem repeat motifs (3228 occurrence; relative frequency=34.24) in GIF, following with trinucleotide tandem repeat motifs (1048 occurrence; relative frequency=11.12). The occurrence of tandem repeats has recorded a decline from tetranucleotide to hexanucleotide. This is expected because the probability of the occurrence of tandem repeats is reduced as the length of motif is increased. Our results are consistent with the findings of Astolfi et al. [47] in general, the work which demonstrated that eukaryotic genes (that of human chromosome 21+22, M. musculus, D. melanogaster, C. elegans, A. thaliana, and S. cerevisiae) are abundant with dinucleotide repeat motif in the comparison with trinucleotide and tetranucleotide repeat motif. Besides, the report of Ouyang et al. [48] was also consistent with our results in terms of the abundance of different tandem repeat motifs, with the exception that mononucleotide tandem repeats were observed being the most abundant in their study (relative frequency=4.36, genome-wide). The deviation between the findings of Ouyang et al. and ours in the abundance level of mononucleotide tandem repeat could be accounted by the fact that different organisms from different kingdoms have been used as the study subject, that is, human gene being the subject in this research whereas virus gene was the subject in Ouyang et al.'s study. However, by comparing the relative frequency of mononucleotide tandem repeats of our findings and that of Ouyang et al. [48], it was found that both our result (relative frequency=2.25) and Ouyang et al.'s (relative frequency=4.36) do not mark a very high value of relative frequency. This implies that the occurrence of mononucleotide tandem repeats in human and virus gene is relatively scarce in the complete gene sequence. Besides, a comparison between the relative frequency of other types of tandem repeat motif shows that the human GIF has more tandem repeats than virus gene.

The number of repeat-in-tandem for the mononucleotide is given in Table 2.

Table 2. Number of repeat-in-tandem for mononucleotide

| Motif with the no. of repeat-in-tandem | No. of occurrence | Percentage |
|---|---|---|
| $A_6$ | 42 | 42% |
| $A_7$ | 16 | 16% |
| $A_8$ | 6 | 6% |
| $A_9$ | 10 | 10% |
| $A_{10}$ | 3 | 3% |
| $A_{11}$ | 2 | 2% |
| $A_{13}$ | 3 | 3% |
| $A_{14}$ | 2 | 2% |
| $A_{15}$ | 3 | 3% |
| $A_{16}$ | 1 | 1% |
| $A_{17}$ | 2 | 2% |
| $A_{18}$ | 5 | 5% |
| $A_{20}$ | 1 | 1% |
| $A_{21}$ | 1 | 1% |
| $A_{23}$ | 1 | 1% |
| $A_{25}$ | 1 | 1% |
| $A_{43}$ | 1 | 1% |
| $T_6$ | 47 | 55.95% |
| $T_7$ | 14 | 16.67% |
| $T_8$ | 12 | 14.29% |
| $T_9$ | 5 | 5.95% |
| $T_{10}$ | 2 | 2.38% |
| $T_{12}$ | 1 | 1.19% |
| $T_{13}$ | 1 | 1.19% |
| $T_{16}$ | 1 | 1.19% |
| $T_{19}$ | 1 | 1.19% |
| $C_6$ | 14 | 93.33% |
| $C_{11}$ | 1 | 6.67% |
| $G_6$ | 12 | 92.31% |
| $G_7$ | 1 | 7.69% |

In Table 2, the number of repeat-in-tandem is analyzed for each mononucleotide motif. G and C tandem repeats do not have a wide variety of repeat length as A and T tandem repeats do. Most of the tandem repeats of C and G, which are more than 90%, have the length of 6 for the repeat-in-tandem. Although a majority of A and T tandem repeats also demonstrates a length of 6 for the repeat-in-tandem, the percentage is apparently lower (mononucleotide $A_6$=42%; $T_6$=55.95%) than that of C (93.33%) and G (92.31%).

The analysis of the number of repeat-in-tandem for the dinucleotide is given in Table 3.

Table 3. Number of repeat-in-tandem for dinucleotide

| Motif with the no. of repeat in tandem | No. of occurrence | Percentage |
|---|---|---|
| $(TC)_2$ | 326 | 91.32% |
| $(TC)_3$ | 27 | 7.56% |
| $(TC)_4$ | 3 | 0.84% |
| $(TC)_5$ | 1 | 0.28% |
| $(CG)_2$ | 61 | 91.04% |
| $(CG)_3$ | 6 | 8.96% |
| $(AG)_2$ | 338 | 93.11% |
| $(AG)_3$ | 20 | 5.51% |
| $(AG)_4$ | 5 | 1.38% |
| $(GC)_2$ | 112 | 90.32% |
| $(GC)_3$ | 12 | 9.68% |
| $(GA)_2$ | 319 | 91.40% |
| $(GA)_3$ | 26 | 7.45% |
| $(GA)_4$ | 3 | 0.86% |
| $(GA)_5$ | 1 | 0.29% |
| $(AT)_2$ | 299 | 90.06% |
| $(AT)_3$ | 26 | 7.83% |
| $(AT)_4$ | 5 | 1.51% |
| $(AT)_6$ | 2 | 0.60% |
| $(AC)_2$ | 181 | 94.76% |
| $(AC)_3$ | 10 | 5.24% |
| $(CA)_2$ | 288 | 88.34% |
| $(CA)_3$ | 35 | 10.74% |
| $(CA)_4$ | 3 | 0.92% |
| $(CT)_2$ | 316 | 90.81% |
| $(CT)_3$ | 29 | 8.33% |
| $(CT)_4$ | 3 | 0.86% |
| $(TG)_2$ | 283 | 91.29% |
| $(TG)_3$ | 18 | 5.81% |
| $(TG)_4$ | 5 | 1.61% |
| $(TG)_5$ | 1 | 0.32% |
| $(TG)_6$ | 1 | 0.32% |
| $(TG)_7$ | 1 | 0.32% |
| $(TG)_{16}$ | 1 | 0.32% |
| $(GT)_2$ | 210 | 85.71% |
| $(GT)_3$ | 28 | 11.43% |
| $(GT)_4$ | 6 | 2.45% |
| $(GT)_{14}$ | 1 | 0.41% |
| $(TA)_2$ | 194 | 89.81% |
| $(TA)_3$ | 17 | 7.87% |
| $(TA)_4$ | 3 | 1.39% |
| $(TA)_6$ | 1 | 0.46% |
| $(TA)_{10}$ | 1 | 0.46% |

Of a total of 12 types of dinucleotide motifs, it is apparent that all of them dominantly exhibit 2 repeats-in-tandem $(XY)_2$. Higher order of repeat-in-tandem $[(XY)_n$, where $n \geq 10]$ has been found in dinucleotide TG, GT, and TA. However, high order of repeat-in-tandem is rare. For the order $n>3$, all 12 types of dinucleotide motifs mark an occurrence of lesser than 3%. Besides, there is a drastic reduced in percentage from the order $n=2$ to $n=3$. For example, $(AG)_2$ marks 93.11% whereas $(AG)_3$ marks 5.51%, which is a gap of 87.6%.

We observed the similar phenomena of the distribution of repeat-in-tandem for tetra-, penta-, and hexanucleotide motifs.

Most of the motifs exhibit an order of 2 of repeat-in-tandem. Table 4 summarizes the repeat-in-tandem for tetra-, penta- and hexanucleotide motif.

Table 4. Summary of the repeat-in-tandem for tetra-, penta-, and hexanucleotide motifs

| Nucleotide motifs with the order of repeat-in-tandem, $n$ | No. of occurrence | Percentage |
|---|---|---|
| Tetranucleotide: | | |
| $n=2$ | 208 | 94.98% |
| $n=3$ | 7 | 3.20% |
| $n>3$ | 4 | 1.83% |
| | | |
| Pentanucleotide: | | |
| $n=2$ | 38 | 95% |
| $n=3$ | 2 | 5% |
| $n>3$ | 0 | 0% |
| | | |
| Hexanucleotide: | | |
| $n=2$ | 8 | 100% |
| $n=3$ | 0 | 0% |
| $n>3$ | 0 | 0% |

From Table 4, it is striking that more than 90% of the nucleotide motifs exhibit $n=2$ for the order of repeat-in-tandem. This result is comparable and consistent with the patterns of dinucleotide motifs shown in Table 3. Pentanucleotide motifs and hexanucleotide motifs do not have any repeat-in-tandem that is larger than an order of 3. To take CCTCTC for example, this hexanucleotide motif occurs twice in GIF gene, both in the form of (CCTCTC)$_2$, which are located at loci 943—954 and 67776—67787, respectively. It was found that these nucleotide motifs have lesser repeat-in-tandem of the order larger than 3. Our finding is partially supported by the research done by Yang et al. [49], who have shown that the genome of *Shigella flexneri* serotype 2a (strain Sf301) has a drastic reduction of tandem-in-repeat for the order which is higher than 3 in mono-, di-, and trinucleotide motifs. In their study, they found that the percentage of tandem-in-repeat for the order higher than 3 is only 6.32% in dinucleotide motifs, but 33.33% in hexanucleotide motifs. However, no comparison is possible between our results and theirs for tetra- and pentanucleotide because they did not disclose the detailed repeat-in-tandem for these motif types. Notably, certain tandem repeat motifs have been discovered in dynamic mutation, such as CCTG [50], TCAT [51], and ATTCT [52]. These tandem repeats which have a high mutation rate are disease inclined. In our study, it was found that CCTG occurs 6 times in the gene of GIF, which is accounted for 2.74% of tetranucleotide total number of occurrence. However, the disease-inclined TCAT and ATTCT were not found in the gene.

Because trinucleotide encodes for protein in the coding region, it has special importance in the study of tandem repeat. A triplet classification system [46] was used to categorize the trinucleotide tandem repeats, as shown in Table 5.

Table 5. Distribution of trinucleotide tandem repeats in GIF

| Type | Repeat motifs (with frequency) | Total |
|---|---|---|
| T1 | AAT(18) ATA(17) TAA(17) ATT(44) TTA(8) TAT(16) | 120 |
| T2 | AAG(43) AGA(33) GAA(35) CTT(19) TTC(15) TCT(19) | 164 |
| T3 | AAC(9) ACA(11) CAA(25) GTT(16) TTG(14) TGT(16) | 91 |
| T4 | ATG(18) TGA(20) GAT(6) CAT(18) ATC(21) TCA(18) | 101 |
| T5 | AGT(7) GTA(11) TAG(8) ACT(17) CTA(7) TAC(0) | 50 |
| T6 | AGG(53) GGA(33) GAG(29) CCT(16) CTC(25) TCC(24) | 180 |
| T7 | AGC(17) GCA(29) CAG(32) GCT(22) CTG(29) TGC(32) | 161 |
| T8 | ACG(0) CGA(4) GAC(2) CGT(0) GTC(2) TCG(3) | 11 |
| T9 | ACC(15) CCA(25) CAC(26) GGT(8) GTG(25) TGG(20) | 119 |
| T10 | GGC(7) GCG(10) CGG(6) GCC(18) | 51 |

| | | | |
|---|---|---|---|
| | CCG(3) | | |
| | CGC(7) | | |

Notably, the trinucleotide tandem repeat motifs in the type T8 are scarce, which contributes 11 occurrences (1.05%) out of a total of 1048 trinucleotide occurrences. However, Ouyang et al. [48] have found that the T8 type of trinucleotide tandem repeat motifs of Herpes simplex virus type 1 is at the moderate level of abundance (20.65% of the total trinucleotide occurrence). The deviation between our results and that of Ouyang et al.'s indicates that there is an enormous difference between the type of amino acids encoded by human and viruses. In a research carried out by Astolfi et al. [47], they have found that the occurrences of ACG(T8), ACT(T5), and CCG(T10) were low in human chromosome 21 and 22, M. musculus, D. melanogaster, C. elegans, A. thaliana, and S. cerevisiae. However, our finding on the abundance level of ACG(0) and CCG(3) were low in their respective type, but ACT(17) is the most abundant trinucleotide tandem repeat motif in type T5.

In our study, AGG is the most abundant trinucleotide tandem repeat motif in T6 and across all types. The similar finding was reported by Astolfi et al. [47], where AGG was the most abundant motif in M. musculus.

From our results, we have identified several potential trinucleotide tandem repeat motifs that are associated with diseases. CAG has occurred 32 times, which is relatively high and it has the risk to induce spinocerebellar ataxia [11]. GAA which has occurred 35 times has been identified to exceed the normal range of repeat number (5-30), a case where it is exposed to Friedreich's ataxia [11]. Lastly, CTG which has a frequency of 29 is exposed to the risk of Huntington's disease-like 2, as the normal range of repeat number is 6-27. The analyses of the repeat consensus and the loci distribution of these three disease-predisposed trinucleotide tandem repeat motifs are given in Table 6.

Table 6. The repeat consensus and the loci distribution of the three disease-predisposed motifs CAG, GAA, and CTG

| Repeat no | Repeat | Start | End |
|---|---|---|---|
| 9 | $(CAG)_2$ | 141 | 146 |
| 66 | $(CAG)_2$ | 1190 | 1195 |
| 274 | $(CAG)_2$ | 5348 | 5353 |
| 284 | $(CAG)_2$ | 5573 | 5578 |
| 296 | $(CAG)_2$ | 5901 | 5906 |
| 457 | $(CAG)_2$ | 9105 | 9110 |
| 467 | $(CAG)_2$ | 9330 | 9335 |
| 479 | $(CAG)_2$ | 9658 | 9663 |
| 598 | $(CAG)_2$ | 11958 | 11963 |
| 623 | $(CAG)_2$ | 12459 | 12464 |
| 643 | $(CAG)_2$ | 13036 | 13041 |
| 676 | $(CAG)_2$ | 13680 | 13685 |
| 710 | $(CAG)_2$ | 14240 | 14245 |
| 776 | $(CAG)_2$ | 15481 | 15486 |
| 1541 | $(CAG)_2$ | 30068 | 30073 |
| 2121 | $(CAG)_2$ | 40507 | 40512 |
| 2255 | $(CAG)_2$ | 43133 | 43138 |
| 2256 | $(CAG)_3$ | 43142 | 43150 |
| 2329 | $(CAG)_2$ | 44418 | 44423 |
| 2552 | $(CAG)_2$ | 48613 | 48618 |
| 2625 | $(CAG)_2$ | 50149 | 50154 |
| 2696 | $(CAG)_2$ | 51699 | 51704 |
| 2706 | $(CAG)_2$ | 51924 | 51929 |
| 154 | $(GAA)_2$ | 2767 | 2772 |
| 251 | $(GAA)_2$ | 4887 | 4892 |
| 262 | $(GAA)_2$ | 5103 | 5108 |
| 282 | $(GAA)_2$ | 5541 | 5546 |
| 291 | $(GAA)_2$ | 5766 | 5771 |
| 431 | $(GAA)_2$ | 8680 | 8685 |
| 442 | $(GAA)_2$ | 8896 | 8901 |
| 465 | $(GAA)_2$ | 9298 | 9303 |
| 474 | $(GAA)_2$ | 9523 | 9528 |
| 732 | $(GAA)_2$ | 14696 | 14701 |
| 940 | $(GAA)_2$ | 18706 | 18711 |
| 1094 | $(GAA)_2$ | 21658 | 21663 |
| 1223 | $(GAA)_2$ | 23986 | 23991 |
| 1321 | $(GAA)_2$ | 25989 | 25994 |
| 1329 | $(GAA)_2$ | 26128 | 26133 |
| 1649 | $(GAA)_2$ | 32116 | 32121 |
| 1657 | $(GAA)_2$ | 32336 | 32341 |
| 1663 | $(GAA)_2$ | 32426 | 32431 |
| 1687 | $(GAA)_2$ | 32851 | 32856 |
| 1872 | $(GAA)_2$ | 35980 | 35985 |
| 2117 | $(GAA)_2$ | 40413 | 40418 |
| 2233 | $(GAA)_2$ | 42742 | 42747 |
| 2567 | $(GAA)_2$ | 48907 | 48912 |
| 2673 | $(GAA)_2$ | 51238 | 51243 |
| 2684 | $(GAA)_2$ | 51454 | 51459 |
| 2704 | $(GAA)_2$ | 51892 | 51897 |
| 2713 | $(GAA)_2$ | 52117 | 52122 |
| 2853 | $(GAA)_2$ | 55031 | 55036 |
| 2864 | $(GAA)_2$ | 55247 | 55252 |
| 2884 | $(GAA)_2$ | 55685 | 55690 |
| 2893 | $(GAA)_2$ | 55910 | 55915 |
| 3089 | $(GAA)_2$ | 59744 | 59749 |
| 3368 | $(GAA)_2$ | 65229 | 65234 |
| 4676 | $(GAA)_2$ | 89872 | 89877 |
| 4710 | $(GAA)_2$ | 90353 | 90358 |
| 21 | $(CTG)_3$ | 323 | 331 |
| 145 | $(CTG)_2$ | 2570 | 2575 |
| 634 | $(CTG)_2$ | 12827 | 12832 |
| 650 | $(CTG)_2$ | 13194 | 13199 |
| 657 | $(CTG)_2$ | 13341 | 13346 |
| 787 | $(CTG)_2$ | 15833 | 15838 |
| 925 | $(CTG)_2$ | 18400 | 18405 |
| 951 | $(CTG)_2$ | 18935 | 18940 |
| 1208 | $(CTG)_2$ | 23680 | 23685 |
| 1234 | $(CTG)_2$ | 24215 | 24220 |
| 1411 | $(CTG)_2$ | 27753 | 27758 |
| 1637 | $(CTG)_3$ | 31889 | 31897 |
| 1661 | $(CTG)_2$ | 32402 | 32407 |
| 1685 | $(CTG)_2$ | 32813 | 32818 |
| 2180 | $(CTG)_2$ | 41879 | 41884 |
| 2192 | $(CTG)_2$ | 42139 | 42144 |
| 2309 | $(CTG)_2$ | 43989 | 43994 |
| 2356 | $(CTG)_2$ | 44794 | 44799 |
| 2460 | $(CTG)_2$ | 46926 | 46931 |
| 2525 | $(CTG)_2$ | 48104 | 48109 |
| 3013 | (CTG)$_2$ | 58384 | 58389 |
| 3163 | (CTG)$_2$ | 61415 | 61420 |
| 3251 | (CTG)$_3$ | 63036 | 63044 |
| 3599 | (CTG)$_2$ | 69662 | 69667 |
| 4103 | (CTG)$_2$ | 79070 | 79075 |
| 4156 | (CTG)$_2$ | 80023 | 80028 |
| 4343 | (CTG)$_2$ | 83521 | 83526 |
| 4371 | (CTG)$_2$ | 83985 | 83990 |
| 4572 | (CTG)$_2$ | 87928 | 87933 |

A graph is provided in Figure 1 to depict the distribution of these disease-predisposed trinucleotide tandem repeat motifs (as shown in Table 6) in the gene.

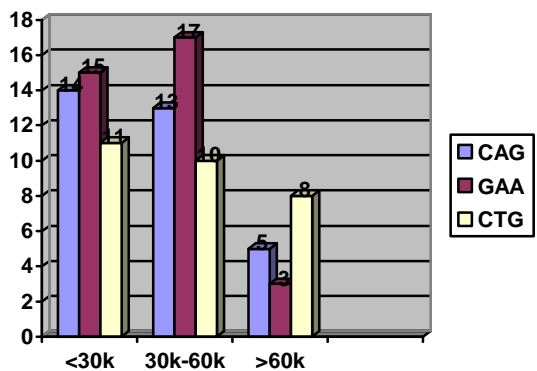

Fig. 1 Distribution of the trinucleotide tandem repeat motifs

From Fig. 1, it is apparent that most of the tandem repeat motifs of these disease-predisposed trinucleotides are located on the first (<30k bp) and the second segment (30k bp – 60k bp) of the gene. For CAG, 14 tandem repeats were found on the first segment whereas 13 tandem repeats were located on the second segment. For GAA, 15 were found on the first segment while 17 motifs were found on the second segment. CTG displays a similar pattern of motif distribution, where 11 were found on the first segment and 10 on the second segment. Therefore, particular attention needs to be paid to these gene segments for therapeutic strategies and drug design.

IV. CONCLUSIONS

An *in silico* analysis has been carried out to study the tandem repeats in GIF. Six types of the nucleotide tandem repeat motifs have been investigated, including mono-, di-, tri-, tetra-, penta- and hexanucleotide. It was found that the relative frequency of dinucleotide tandem repeat motifs is highest, following with that of trinucleotide. The probability of the occurrence of tandem repeats is reduced as the length of motif is increased. Besides, from the analysis it was found that most of the motifs exhibit an order of 2 of repeat-in-tandem, and higher order of repeat-in-tandem is rare in GIF. The analysis of the trinucleotide tandem repeat motifs has identified three motifs which are disease-predisposed. The loci of these motifs have been identified and this information is important for the mutation analysis carried out by geneticists and biomedical scientists. Our analysis of the tandem repeats in GIF sheds light on the better understanding of the diseases implicated by this gene at the molecular level.